\documentclass[%
 aip,a4paper,onecolumn,
 amsmath,amssymb,
reprint
]{revtex4-1}
\usepackage{graphicx,xcolor}
\usepackage{dcolumn}
\usepackage{bm}

\usepackage[T1]{fontenc}
\usepackage{mathrsfs,amssymb,graphicx,epsfig}
\def\rr{}
\renewcommand{\[}{\begin{equation}}
\renewcommand{\]}{\end{equation}}
\def\bea{\begin{eqnarray}}
\def\eea{\end{eqnarray}}
\def\nn{\nonumber\\}
\newcommand{\intr}{\int d{\bf r} \;}

\newcommand{\B}{{\bf B}}

\newcommand{\A}{{\bf A}}

\newcommand{\p}{{\bf p}}

\renewcommand{\k}{{\bf k}}

\newcommand{\kk}{{\mbox{\boldmath$\kappa$}}}
\def\EEE{\mbox{\boldmath${\cal E}$}}
\def\LLL{\mbox{\boldmath${\cal L}$}}
\def\LL{{\cal L}}
\def\EE{{\cal E}}
\def\ZZ{{\cal Z}}
\renewcommand{\v}{{\bf v}}
\renewcommand{\r}{{\bf r}}
\newcommand{\R}{{\bf R}}
\renewcommand{\P}{{\bf P}}

\newcommand{\equ}[1]{Eq.~(\ref{#1})}
\newcommand{\eqs}[2]{Eqs.~(\ref{#1}) and (\ref{#2})}

\def\tensor#1{\stackrel{\leftrightarrow}{#1}}

\def\ket#1{\vert#1\rangle}
\def\ev#1{\langle#1\rangle}
\def\me#1#2#3{\langle#1| \, #2 \, |#3\rangle}
\def\runtime{(\the\time)\qquad\the\month/\the\day/\the\year}
\def\today
 {\count10=\year\advance\count10 by -2000 \number\day--\ifcase
  \month \or Jan\or Feb\or Mar\or Apr\or May\or Jun\or
             Jul\or Aug\or Sep\or Oct\or Nov\or Dec\fi--\number\count10}

\def\hour{\count10=\time\count11=\count10
\divide\count10 by 60 \count12=\count10
\multiply\count12 by 60 \advance\count11 by -\count12\count12=0
\number\count10 :\ifnum\count11 < 10 \number\count12\fi\number\count11}
\begin{document}

\title{Quantum geometry and adiabaticity in molecules and in condensed matter}

\date{\sf DRAFT: run through \LaTeX\ on \today\ at \hour}
\author{Raffaele Resta}
 \email{resta@iom.cnr.it}
\affiliation{ 
CNR-IOM Istituto Officina dei Materiali, Trieste, Italy
}%

 \homepage{http://www-dft.ts.infn.it/\~{}resta/.}
\affiliation{Donostia International Physics Center, 20018 San Sebasti{\'a}n, Spain
}%

\begin{abstract} 
The adiabatic theorem states that when the time evolution of the Hamiltonian is ``infinitely slow'', a system, when started in the ground state, remains in the instantaneous ground state at all times. This, however, does not mean that the adiabatic evolution of a generic observable obtains simply as its expectation value over the instantaneous eigenstate. As a general principle there is an additional adiabatic term, of quantum-geometrical nature, which is the relevant one for several static or adiabatic observables. This is shown explicitly for the  cases of polarizability and infrared tensors (in molecules and condensed matter); rotational $g$ factor and magnetizability (in molecules only). Quantum geometry allows for a transparent derivation and a compact expression for these observables, alternative to the well known sum-over-states Kubo formulas.

\end{abstract}

\maketitle


\section{Introduction}

The quantity which today is commonly addressed as a Berry curvature was introduced in the 1980s by Michael Berry \cite{Berry84} and, independently, by David Thouless and coworkers in a series of papers. The many-body formulation---upon which the present work is rooted---owes to a milestone paper by Niu and Thouless,\cite{niu84} hereafter quoted as NT. These papers established the foundations of what nowadays we call ``quantum geometry''.

The term ``adiabatic'' intuitively refers to the ``infinitely slow''  time evolution of a physical system. In thermodynamics the system  transforms through a series of continuously evolving equilibrium states; in quantum mechanics through a series of continuously evolving ground states.\cite{Kato50} I am addressing here the adiabatic response of some observables in condensed matter and in  molecules, showing that quantum geometry---and in particular the Berry curvature---allows for a compact and elegant formulation of the corresponding tensors in both cases. All these tensors  are a measure of how the many-electron system  responds to a slow evolution of the Hamiltonian, and this is precisely what the Berry curvature allows casting in a very compact formalism.

In condensed-matter physics the role of quantum geometry could be hardly underestimated.\cite{rap_a20,Vanderbilt,Xiao10}  In molecular physics, instead, quantum geometry is endorsed since long time to address conical intersections and related problems,\cite{Mead92} while  its role in the formulation of the adiabatic-response tensors went unnoticed until recently.\cite{rap168} 

The first part of this work is formulated in a generic quantum mechanical setting, i.e. not specifically addressing electrons.
 In Sec. \ref{sec:qg} I provide some  fundamentals of quantum geometry; in Sec. \ref{sec:adia} I give a sharp definition of what ``infinitely slow'' means, in the sense of Kato's theorem;\cite{Kato50} in Sec. \ref{sec:hf} it is shown that the adiabatic evolution of an observable can be expressed in a very compact form, referred to in the following as the ``adiabatic Hellmann-Feynman theorem'', \equ{HF} below. The two terms therein can be regarded as the nongeometrical and geometrical one, respectively.
 
 A generic many-electron system, either condensed-matter or molecular, is addressed since Sec. \ref{sec:many} onwards. The explicit expressions for the chosen observables are presented in parallel, using---to the extent of possible---a common formulation and common notations for both molecules and condensed matter. Some notable differences, whenever occurring, are emphasized. 
 The response properties addressed in the following are: polarizability and infrared tensors (in molecules and condensed matter); rotational $g$ factor and magnetizability (in molecules only). All these tensor properties have a time-honored expression provided by linear-response theory (i.e. Kubo formulas). It is shown here that they admit an alternative geometrical derivation: they are indeed simple  case studies of  the adiabatic Hellmann-Feynman theorem at the core of the present work.
The expressions  in terms of the appropriate Berry curvature are all very compact and transparent.

\section{Quantum geometry} \label{sec:qg}

Let $\hat{H}$ be a time-independent Hamiltonian and let $\ket{\Psi_n}$ be its eigenstates with eigenvalues $E_n$; when  $\hat{H}$ depends on a generic real parameter $\lambda$,  the eigenstates and eigenvalues are parameter-dependent as well. The ground state is assumed to be nondegenerate for all $\lambda$; its Berry connection is then defined as \[ {\cal A}_\lambda = i \ev{\Psi_0|\partial_\lambda \Psi_0} , \]
and measures the infinitesimal phase variation of the eigenstate when $\lambda$ is varied: \[ d\phi = - \mbox{Im ln} \ev{\Psi_{0\lambda} | \Psi_{0\lambda+d\lambda} } = {\cal A}_\lambda d \lambda.  \label{berry} \] Since the eigenstate $\ket{\Psi_{0\lambda}}$ is arbitrary by a $\lambda$-dependent phase factor, the connection is gauge-dependent and cannot be endowed---as such---with any physical meaning; because of this its role has been overlooked for many years.

In a pathbreaking paper \cite{Berry84}  Michael Berry realized that, whenever the Hamiltonian is periodical in $\lambda$, the loop integral of the connection \[ \gamma = \oint d \phi \] is a gauge-invariant phase---called nowadays Berry phase---defined modulo $2\pi$. The basic tenet of Berry's paper is that any gauge-invariant quantum-mechanical expression corresponds in principle to an observable. Since then, several observables having the nature of a Berry phase have been identified in electronic structure theory \cite{rap_a20,Xiao10,Vanderbilt} and in many other domains. 

Next let us suppose that $\hat{H}$ also depends on a second real parameter $\kappa$;
the ground state is assumed to be nondegenerate for all $(\kappa,\lambda)$. 
The Berry curvature is by definition
\[ \Omega(\kappa,\lambda) =  \partial_\kappa {\cal A}_\lambda - \partial_\lambda {\cal A}_\kappa   = - 2 \,\mbox{Im } \ev{\partial_\kappa \Psi_0 | \partial_\lambda \Psi_0} , \label{curva}\] 
and is gauge-invariant. Let $\Sigma$ be a domain in the $\kappa\lambda$ plane where $\Omega(\kappa,\lambda)$ is regular, and let $\partial \Sigma$ be its boundary.; then Stokes' theorem insures that \[ \int_\Sigma d\kappa d\lambda \; \Omega(\kappa,\lambda) = \gamma, \] where $\gamma$ is the Berry phase \[ \gamma = \oint_{\partial \Sigma} ( {\cal A}_\lambda d \lambda +  {\cal A}_\kappa d\kappa) . \] Therefore the Berry curvature has the meaning of Berry phase per unit area.

The parameters may have various physical interpretations and different dimensions; the curvature has the inverse dimensions of the product $\kappa \lambda$. For a macroscopic homogeneous system $\Omega(\kappa,\lambda)$ is an extensive quantity.

When $(\kappa,\lambda)$ becomes time-dependent, the curvature
is a quasi-static quantity, in the sense that its definition requires solely the instantaneous ground eigenstate; yet it encodes the lowest-order effect of the excited states on the adiabatic evolution. 

In the parallel-transport gauge\cite{Vanderbilt} the $\kappa$-derivative of the eigenstate can be written as \[ \ket{\partial_\kappa \Psi_0} = \sum_{n\neq 0} \ket{\Psi_n} \frac{\me{\Psi_n}{\partial_\kappa \hat{H}}{\Psi_0}}{E_0 - E_n} \label{parallel} , \] and analogously for the $\lambda$-derivative. Therefore the curvature admits a sum-over-states (i.e. Kubo) formula, first displayed in the original Berry's paper:
\[ \Omega(\kappa,\lambda) = -2 \, \mbox{Im}  \sum_{n \neq 0} \frac{\ev{\Psi_0 | \partial_\kappa \hat{H} | \Psi_n }\ev{\Psi_n | \partial_\lambda \hat{H} | \Psi_0 }}{(E_0 - E_n)^2} . \label{sum} \]  The formula perspicuously shows how the Berry curvature encodes---to lowest order---the effect of the excited states on the ground state when the quantum system is transported in the parameter space. \equ{sum}
has also the virtue of showing that the curvature becomes ill defined whenever the ground state is degenerate with the first excited state; for instance, at a conical intersection. 

\section{What does it mean ``adiabatic''?} \label{sec:adia}

Intuitively, the concept of adiabaticity is related to the time evolution of a system driven out of equilibrium ``infinitely slowly''; in quantum mechanics, the Hamiltonian is assumed to have an ``infinitely slow'' time dependence.\cite{Kato50} In the current literature the terms ``adiabatic'' and ``nonadiabatic'' are sometimes improperly used; it is therefore expedient to start with a sharp definition. 

Suppose we have a time-independent Hamiltonian which depends on a parameter $\lambda$. When this parameter acquires a time dependence, $\lambda \rightarrow \lambda(t)$, the system and its observables evolve in time; saying that such evolution is ``infinitely slow'' is equivalent to say that $\dot\lambda(t)$ is infinitesimal.

If $\hat{O}$ is a time-independent operator, the exact time-evolution of the corresponding observable is \[ O(t) \equiv \me{\Psi_t}{\hat{O}}{\Psi_t}, \label{exact} \]  where $\ket{\Psi_t}$ is the solution of the time-dependent Schr\"odinger equation with $\hat{H}_t = \hat{H}_{\lambda(t)}$. The adiabatic evolution of an observable, as addressed throughout this work, coincides with the exact evolution up to linear order in $\dot\lambda(t)$ and neglects terms of order $\ddot\lambda(t)$ and $\dot\lambda^2(t)$ (and higher). Equivalently, if the time evolution of the Hamiltonian is harmonic at frequency $\omega$, the adiabatic evolution of $O(t)$ coincides with the exact one for infinitesimal $\omega$: i.e. to linear order in $\omega$, neglecting terms of order $\omega^2$ and higher. As shown below, the adiabatic evolution of a generic observable obtains by solving the {\it time-independent} Schr\"odinger equation at $\lambda = \lambda(t)$.

In the chemical-physics literature, adiabaticity has been generally dealt with in the context of the Born-Oppenheimer (also called Born-Huang) decoupling of the equations of motion for a system of electrons and nuclei.\cite{Nafie83,Patchkovskii12,Hanasaki21} Here, instead, I address adiabaticity per se, keeping the issue as conceptually distinct: the Born-Oppenheimer approximation is---as the name says---an approximation, while the adiabatic limit enjoys an exact definition.  Furthermore I am addressing here  the more general case where  the time dependence of the Hamiltonian does not {\rr necessarily} owe to the nuclear motion.

\section{The adiabatic evolution of an observable} \label{sec:hf}

The adiabatic theorem states that\cite{Kato50} ``when the change of the Hamiltonian in time is made infinitely slow, the system, when started from a stationary state, passes through the corresponding stationary states for all times''. It is therefore tempting---in order to express the adiabatic evolution of an observable---to replace the exact state vector $\ket{\Psi_t}$ in \equ{exact} with the ground state $\ket{\Psi_0}$ evaluated at $\lambda(t)$. This is obviously incorrect: in a time-reversal invariant system the eigenfunctions can be taken as real. If $\hat{O}$ is an imaginary operator---like the current or the angular momentum---the expectation value $\me{\Psi_0}{\hat{O}}{\Psi_0}$ vanishes at all times, while there is no reason for $O(t)$ to vanish. The correct expression can be derived in a very transparent way in the special case where the operator $\hat{O}$ is the derivative of the Hamiltonian $\hat{H}$ with respect to some parameter, as shown next.

\subsection{The adiabatic Hellmann-Feynman theorem}

Let us assume that a generic time-independent operator $\hat{O}$ can be written as the derivative of the Hamiltonian with respect to the parameter $\kappa$: \[ \hat{O} = \partial_{\kappa}  \hat{H}_t. \]  It will be found below that several interesting observables belong to this class. For a time-independent $\lambda$ the Hellmann-Feynman theorem yields \[ O = \me{\Psi_0}{\hat{O}}{\Psi_0}= \partial_\kappa E_0 . \]
When instead $\lambda$ depends on time the $\kappa$-derivative of the energy is \bea \partial_\kappa E_t &=& \partial_\kappa  \me{\Psi_t}{\hat{H}_t}{\Psi_t} \nn &=& \me{\Psi_t}{\hat{O}}{\Psi_t}  +  \me{\partial_\kappa \Psi_t}{\hat{H}_t}{\Psi_t}  -  \me{\Psi_t}{\hat{H}_t}{\partial_\kappa\Psi_t} \nn &=& \me{\Psi_t}{\hat{O}}{\Psi_t} + i\hbar \, (\, \ev{\partial_{\kappa}\Psi_t |\dot\Psi_t} - \ev{\dot\Psi_t |\partial_{\kappa}\Psi_t} \,) . \eea Therefore the {\it exact} evolution of the observable is \[ O(t) \equiv \me{\Psi_t}{\hat{O}}{\Psi_t}  = \partial_{\kappa} E_t + 2 \, \hbar \,\mbox{Im } \ev{\partial_{\kappa}\Psi_t|\dot\Psi_t}  . \label{exact2} \]  

At this point we may exploit the adiabatic theorem:\cite{Kato50} if $\ket{\Psi_t} =  \ket{\Psi_0}$ at $t=0$, and if the time evolution of $\lambda(t)$ is ``infinitely slow'',  one may replace  $E_t$ and $\ket{\Psi_t}$ with their instantaneous value, {\rr and $\ket{\dot\Psi_t}$ with  $\ket{\partial_\lambda\Psi_0}\dot\lambda(t)$, thus}
obtaining  the adiabatic limit as \bea O(t) &=& \partial_{\kappa} E_0 - \hbar \, \Omega(\kappa,\lambda)\, \dot\lambda(t) \nn  &=& {\rr
\me{\Psi_0}{\hat{O}}{\Psi_0}  - \hbar \, \Omega(\kappa,\lambda)\, \dot\lambda(t) }, \label{HF} \eea where $\ket{\Psi_0}$, $E_0$, and the curvature depend implicitly on time. This expression will be referred throughout this paper as the adiabatic Hellmann-Feynman theorem; to the best of the author's knowledge it first appeared in the NT paper.

\subsection{A generic observable}

Let us consider next the more general case where the time-independent operator $\hat O$ {\it cannot} be expressed as an Hamiltonian derivative. The expression for  $O(t)$ in the adiabatic limit is {\rr instead provided by}  the sum-over-states form of the Berry curvature, \equ{sum}: \bea {O(t)} &=& \me{\Psi_0}{\hat{O}}{\Psi_0} \nn &+& 2 \hbar \dot\lambda(t) \, \mbox{Im}  \sum_{n \neq 0} \frac{\ev{\Psi_0 | \hat{O} | \Psi_n }\ev{\Psi_n | \partial_\lambda \hat{H} | \Psi_0 }}{(E_0 - E_n)^2} , \label{sum2} \eea where once more the eigenstates and the eigenvalues are evaluated at $\lambda(t)$. Indeed, \eqs{sum2}{HF} are both proved in the  NT paper,
where the exact time evolution of $O(t)$ is expanded to the order $\dot\lambda(t)$, explicitly neglecting terms of order $\ddot\lambda(t)$ and $\dot\lambda^2(t)$. {\rr  Actually \equ{HF} was  originally obtained  by NT as a special case of \equ{sum2}}.

\subsection{Time-reversal invariance}

The previous major results, \eqs{HF}{sum2}, show that the adiabatic evolution of an observable is comprised in general of two terms, which one might define the nongeometrical and geometrical one, respectively; the expressions nonetheless simplify in presence of time-reversal (T) symmetry, as in the case studies addressed below.
In a T-symmetric system all unperturbed eigenfunctions can be chosen as real. Therefore whenever $\hat{O}$ is a real operator (like the density) only the first term in  \eqs{HF}{sum2} is nonvanishing; conversely whenever $\hat{O}$ is an imaginary operator (like a current or an angular momentum) only the second  term in  \eqs{HF}{sum2} is nonvanishing. In the latter case the adiabatic HF theorem  takes the simple, purely geometrical,  form \[ \frac{O(t)}{\dot\lambda(t)} = - \hbar \, \Omega(\kappa,\lambda), \qquad \mbox{$\hat{O}$ imaginary}  \label{key} . \] 

\section{Many-electron systems} \label{sec:many}

All of the above results are very general: they address any quantum system, whose Hamiltonian depends on two arbitrary parameters. From now on we consider the electronic Hamiltonian of a system of $N$ interacting electrons in the Born-Oppenheimer {\rr approximation} (sometimes called Born-Huang): \[ \hat{H}  = \frac{1}{2m} \sum_{i=1}^N \left({\bf p}_i + \frac{e}{c} \A(\r_i) \right)^2 + \hat{V} ,\label{H}\]  where the potential $\hat{V}$ comprises one-body and two-body terms.
In the following we will consider cases where $\A(\r)$ depends on some parameter $\kappa$, and $\hat{V}$ depends on some parameter $\lambda$. $\hat{V}$  is a multiplicative operator in the Schr\"odinger representation; it is a function of the instantaneous nuclear positions $\{\R_s\}$, and possibly of other parameters.  We assume a singlet ground state and we neglect irrelevant spin variables. 
In the following of this work Greek subscripts are Cartesian indices, and the sum over repeated indices is implicitly understood. Gaussian units are adopted throughout; the Hamiltonian is T-invariant at $\A=0$.

The eigenvalue problem is defined once the boundary conditions are imposed to the solutions of  the Schr\"odinger equation. These are profoundly different in the two cases dealt with in this work: molecules and condensed matter. A given choice establishes the Hilbert space to which the state vectors belong.

When addressing the bounded states of a molecule one imposes  ${\mathscr L}^2$ boundary conditions:  the wavefunction vanishes far away from the system. In the condensed-matter theory jargon such boundary conditions are defined ``open'' (OBCs).

In the condensed-matter case one addresses instead an unbounded many-electron system within Born-von-K\`arm\`an  periodic boundary conditions (PBCs): the many-body wavefunctions are periodic with period $L$ over each electronic Cartesian coordinate $r_{i\alpha}$ independently.
One considers therefore a system of $N$ interacting electrons in a cubic box---often called ``supercell''---of volume $L^3$, together with its periodic replicas. The potential $\hat{V}$ obeys PBCs and includes the classical nuclear-nuclear repulsion; the supercell is charge-neutral.
The limit $N \rightarrow \infty$,  $L \rightarrow \infty$, $n = N/L^3 $ constant is understood. It is not required that the system is crystalline, only that it is macroscopically homogeneous; owing to PBCs the macroscopic field $\EEE$ is set to zero.

While in insulators adiabaticity is obvious (the spectrum is gapped), the metallic case---also dealt with in the following---requires a crucial observation. Following a milestone 1964 paper by Kohn,\cite{Kohn64} $\kappa$- and $\lambda$-derivatives will be evaluated first, and the $L \rightarrow \infty$ limit taken afterwards. At any finite size the spectrum is gapped even in metals; the procedure warrants that even the response of a metallic system to an infinitely slow perturbation is indeed adiabatic.

Equivalently, one may consider time scales. The inverse of the finite-$L$ gap sets the internal time scale of the many-electron system, while the Hamiltonian evolves in time $t$. The adiabatic limit is the $t \rightarrow \infty$ limit, and such limit is here performed {\it before} the  $L \rightarrow \infty$ one.

\section{Some observables} \label{sec:obs}

In the following I will address a few electronic observables, by adopting a common setting in both the molecular case and in the condensed-matter case. It is expedient to provide the explicit expression of the operators which define such observables.

I start with a peculiar observable; the dipole, i e. the position operator \[ \hat{\r} = \sum_{i=1}^N \r_i . \] This operator is a trivial multiplicative operator  only within OBCs; it is instead a {\it forbidden}  operator within PBCs.\cite{rap100} In fact, a Born-von-K\`arm\`an  periodic wavefunction multiplied by $ \hat{\r}$ does not belong to the PBCs Hilbert space. Nonetheless the off-diagonal elements of $\hat{\r}$ are usually expressed in terms of the many-body {\rr velocity
\[ \hat\v = \frac{1}{\hbar} [ \hat{H} , \hat\r ]  . \label{momentum}  \] 
In the rest of this paper only T-symmetric systems will be addressed, where the vector potential vanishes at equilibrium; therefore $\hat\v = \hat\p/m$ and 
}\[ \me{\Psi_n}{\hat{\r}}{\Psi_{n'}} = i \frac{\hbar}{m} \frac{ \me{\Psi_n}{\hat\p}{\Psi_{n'}}}{E_{n'} - E_n} ; \qquad n\neq n'  . \label{comm} \] It is worth stressing  that \equ{comm} is an identity within OBCs, while it is a {\it definition} within PBCs.

Operators which will be addressed below are: 
\begin{itemize}
\item Charge density:
\vspace{-0.5cm}\[ \hat\rho(\r)= -e  \sum_{i=1}^N \delta(\r-\r_i).  \label{rho} \]
\item Current density when ${\bf A}(\r)=0$:
\vspace{-0.3cm}\[ \hat{\bf j}(\r) = - \frac{e}{2m} \sum_{i=1}^N [ \delta(\r-r_i)\p_i + \p_i \delta(\r-r_i) ] \label{currdef}  .\]
\item Integrated current (extensive):
\[ \hat{{\bf J}} = \int d\r \; \hat{\bf j}(\r) = - \frac{e}{m} \hat \p=  -c \, \partial_{{\bf A}} \hat{H}.  \label{integrated}\]
Notably, the integrated-current operator can be expressed as the Hamiltonian derivative with respect to ${\bf A}$, evaluated at ${\bf A} = 0$.
\item Angular momentum (in molecules only):
\[ {\rr \hat{\LLL} } =  \sum_i \r_i\times \p_i . \label{pva} \]
\end{itemize}

\section{Continuity}

The most fundamental adiabatic response is the microscopic electrical current density, which explicitly displays how the electronic charge is ``dragged'' by a slow variation of the Hamiltonian.
It is by definition the expectation value of the operator $\hat{\bf j}(\r)$; the quantum-chemical literature equivalently addresses the electronic flux,\cite{Nafie83,Patchkovskii12,Hanasaki21,rap166} whose operator differs from \equ{currdef} by the $-e$ factor. 

Continuity equation is an exact relationship between the time derivative of the charge density and the divergence of the current density: \[ \dot\rho(\r,t) = - \nabla \cdot {\bf j}(\r,t) . \label{continuity} \] It is shown in the following that \equ{continuity} is conserved when both members are evaluated in the adiabatic limit.

The operator $\hat{\bf j}(\r)$ is imaginary, ergo only the second term of \equ{sum2} contributes to the adiabatic evolution: \[ {\bf j}(r,t) = 2 \hbar \dot\lambda(t) \, \mbox{Im}  \sum_{n \neq 0} \frac{\ev{\Psi_0 | \hat{{\bf j}}(\r) | \Psi_n }\ev{\Psi_n | \partial_\lambda \hat{H} | \Psi_0 }}{(E_0 - E_n)^2} . \label{nafie} \]  Expressions for the adiabatic current density---equivalent to this one---first appeared in molecular physics in the 1980s, quite independently from the NT result.\cite{Nafie83}

Since $\hat\rho(\r)$ is a real operator, only the first term in \equ{sum2} contributes:
 \[ \dot\rho(\r,t)= 2 \dot\lambda(t) \, \mbox{Re }\me{\Psi_0}{\hat\rho(\r)}{\partial_{\lambda}\Psi_0}  . \] What remains  to be proved is  that \bea &&  \mbox{Re }  \me{\Psi_0}{\hat{\rho}(\r)}{\partial_\lambda\Psi_0} \nn &= &\!\!\!  -  \hbar \, \mbox{Im}  \sum_{n \neq 0} \frac{\ev{\Psi_0 | \nabla \cdot \hat{\bf j}(\r) | \Psi_n }\ev{\Psi_n | \partial_\lambda \hat{H} | \Psi_0 }}{(E_0 - E_n)^2} ; \eea this is straightforward by making use of the identity\cite{rap166}
 \[ \nabla \cdot \hat{{\bf j}}(\r)= \frac{i}{\hbar}  [ \,\hat\rho(\r) , \hat{H} \,] . \label{rapix} \]
It is worth stressing that  \equ{rapix} only holds for a multiplicative potential $\hat{V}$; nonlocal pseudopotentials violate continuity.

\section{Macroscopic fields} \label{sec:fields}

Since throughout this work I am addressing both molecules and condensed matter, it is worth stressing an important difference between the two cases in what concerns the macroscopic electric field.

In the molecular case the experimenter applies an {\it external} electric field $\EEE_0$, by means e.g. of a capacitor. The microscopic screened field $\EEE(\r)$ equals the applied field plus the induced field. The latter decays polynomially far form the molecule (induced multipoles), hence the field $\EEE$ coincides with $\EEE_0$ far from the molecule. This statement is not as innocent as it seems. In fact a given charge distribution $\rho(\r)$ does not determine the field  $\EEE(\r)$ uniquely: the solution of Poisson's equation is defined only modulo a solution of the corresponding homogeneous equation. Obviously $\EEE(\r) =$ constant is a solution of the homogeneous equation: therefore  setting  $\EEE = \EEE_0$ is actually a tacit choice of the integration constant.

In the condensed-matter case, instead, the many-electron system is unbounded and there is no ``external''  field $\EEE_0$ to speak of. On a finite sample  the experimenter directly controls (by means e.g. again of a capacitor)
the {\it screened} field $\EEE$ The latter is  defined as the macroscopic  average of the microscopic one  $\EEE(\r)$.\cite{Jackson} 
As said above setting the value of $\EEE$ amounts to a choice of boundary conditions in integrating Poisson equation.  In {\rr the great majority of}  selfconsistent calculations the Coulomb potential is taken as lattice-periodical, ergo the condition $\EEE=0$ is imposed; different choices are possible, though.\cite{Souza02,Umari02}

\section{Linear-response tensors}

\subsection{Polarizability}

I deal here with the electronic term only in the linear response, also called ``clamped nuclei'' response. The molecular and condensed-matter polarizability tensors are defined as \[ \tensor\alpha = \frac{\partial {\bf d}}{\partial \EEE} \quad \mbox{and} \quad \tensor\chi = \frac{\partial \P}{\partial \EEE} , \label{tensors} \] respectively, where ${\bf d}$ is the molecular dipole and $\P$ is the macroscopic polarization; the dielectric tensor $\tensor\varepsilon_\infty$  is defined  as\cite{Kittel} \[ \tensor\varepsilon_\infty = 1 + 4\pi \tensor\chi  .\] \equ{tensors} defines two static properties. Alternatively, we may assume that the field is adiabatically switched on in time; the tensors admit therefore the equivalent definition \[ \tensor\alpha = \frac{\partial {\bf d}/\partial t}{\partial \EEE/\partial t} ,\qquad \tensor\chi = \frac{\partial \P/\partial t}{\partial \EEE/\partial t} . \label{tensors2} \] 

The time-derivative of ${\bf d}$ coincides with the expectation value of $\hat{\bf J}$, \equ{integrated}, while the time-derivative of $\P$ coincides with the expectation value of $\hat{\bf J}/L^3$; therefore \[ \alpha_{\beta\gamma} =\frac{J_\beta(t)}{\dot\EE_\gamma}, \qquad  \chi_{\beta\gamma} = \frac{1}{L^3} \frac{J_\beta(t)}{\dot\EE_\gamma} . \] In order to exploit the adiabatic HF theorem in the form of \equ{key} we have to identify the parameter $\kappa$ and $\lambda$ entering this case study. Given that $\hat{{\bf J}} =  -c \, \partial_{{\bf A}} \hat{H} $, the parameter $\kappa$ is a Cartesian component of the vector potential $\A$ (taken as $\r$-independent); the Hamiltonian acquires its adiabatic time dependence via the macroscopic field $\EEE$, ergo the parameter $\lambda$ is identified here as one of its Cartesian components. The geometrical expression for the response tensors is therefore
 \[ \alpha_{\beta\gamma} =\hbar c  \, \Omega(A_\beta,\EE_\gamma), \qquad  \chi_{\beta\gamma} = \frac{\hbar c }{L^3}  \Omega(A_\beta,\EE_\gamma), \label{imme} \] where the curvature is evaluated at $\A=0$ and $\EEE=0$. {\rr It is worth mentioning that in the condensed-matter literature the vector potential  is often cast as $\A= \hbar c \kk/e$, where $\kk$---called ``flux'' or ``twist''---has the dimensions of an inverse length;\cite{Xiao10,Kohn64,rap157,rap165} at the independent-particle level $\kk$ is identified with the Bloch vector $\k$.}

The above compact quantum-geometrical expressions are indeed equivalent to the customary Kubo formulas; this is proved in the following. \equ{sum} yields \[  \Omega(A_\beta,\EE_\gamma), = -2 \, \mbox{Im}  \sum_{n \neq 0} \frac{\ev{\Psi_0 | \partial_{A_\beta} \hat{H} | \Psi_n }\ev{\Psi_n | \partial_{\EE_\gamma}\hat{H} | \Psi_0 }}{(E_0 - E_n)^2} . \label{sump} \] We then exploit  $\partial_{A_\beta}\hat{H} = e\hat{p}_\beta/(mc) $, and $\partial_{\EE_\gamma} \hat{H} = e\, \hat{r}_\gamma$ to obtain \bea  \Omega(A_\beta,\EE_\gamma) &=& -\frac{2e^2}{mc} \, \mbox{Im}  \sum_{n \neq 0} \frac{\ev{\Psi_0 | \hat{p}_\beta | \Psi_n }\ev{\Psi_n | \hat{r}_\gamma | \Psi_0 }}{(E_0 - E_n)^2} \nn &=& \frac{2e^2}{\hbar c}  \sum_{n \neq 0} \frac{\ev{\Psi_0 | \hat{r}_\beta | \Psi_n }\ev{\Psi_n | \hat{r}_\gamma | \Psi_0 }}{E_n - E_0} , \label{psum}
\eea where \equ{parallel} has been used to get the second line. Replacement of \equ{psum} into \equ{imme} completes the proof. 

{\rr \eqs{sump}{psum} are formally identical in molecules and in condensed matter, yet the symbols therein indicate different quantities: the eigenstates obey different boundary conditions, and the field $\EEE$ is not the same field, as discussed at the beginning of Sec. \ref{sec:fields}.
} Suppose  one evaluates the polarizability $\tensor\alpha$ of a macroscopic crystallite in the same way as for a (very large) molecule. A glance at \equ{tensors} would suggest that $\tensor\chi$ would be equal to $\tensor\alpha$ divided by the sample volume: this is grossly incorrect and the two quantities are in general different. In fact $\tensor\alpha$ is the response to the {\it external} field ($\EEE$ is identified with $\EEE_0$) while $\tensor\chi$ is the response to the {\it internal} field $\EEE$. The relationship between  $\EEE$ and $\EEE_0$ depends on the crystallite's shape, while $\tensor\chi$ is a bulk property (i.e. shape-independent). More details can be found e.g. in Ref. \onlinecite{rap_a30}. 

\subsection{Infrared-charge tensors} \label{sec:aat}

The atomic infrared-charge tensors are the basic entries to determine the oscillator strength of vibrational modes in the harmonic regime. They are called atomic polar tensors in molecular physics and Born effective-charge tensors in condensed matter physics.

Let  $\R_s$ be the equilibrium nuclear position of nucleus $s$; then
 the (dimensionless) $s$-th infrared tensor  is defined as \[ \ZZ^*_{s,\alpha\beta} =  \frac{1}{e} \frac{\partial {d_\alpha}}{\partial R_{s\beta}} \quad \mbox{and} \quad  Z^*_{s,\alpha\beta}  =  \frac{L^3}{e} \frac{\partial {P_\alpha}}{\partial R_{s\beta}}  , \label{born} \] in molecules and condensed matter, respectively. Both ${\bf d}$ and $\P$ include the nuclear contributions; the systems are charge-neutral. 
 
 As in the previous case, we may equivalently redefine these tensors by means of  adiabatic nuclear displacements: 
 \bea \ZZ^*_{s,\alpha\beta} &=&  \frac{1}{e}  \frac{\partial {d_\alpha /\partial t }}{\partial R_{s\beta} \partial t} =  Z_s \delta_{\alpha\beta} +   \frac{1}{e}  \frac{J_\alpha(t)}{\dot{R}_{s\beta}}  \nn Z^*_{s,\alpha\beta}  &=&  \frac{L^3}{e}  \frac{\partial {P_\alpha /\partial t }}{\partial R_{s\beta} \partial t}   =   Z_s \delta_{\alpha\beta} +   \frac{1}{e}  \frac{J_\alpha(t)}{\dot{R}_{s\beta}} ,  \label{born2} \eea where $Z_s$ is the atomic number of nucleus $s$. {\rr Notice that while formal polarization $\P$ is ill defined in metals,\cite{Vanderbilt}  the second line applies as it stands to both insulators and metals,\cite{Dreyer22,rap165} because a polarization derivative is in fact a current.}
   
It is worth reminding  that the electronic response is adiabatic even in metals, despite the absence of a spectral gap therein. In insulators the Born effective charges manifest themselves in the phonon spectra: they generate a depolarization field $\EEE$ which contributes to the restoring force at the zone center. \cite{Huang50,Cochran62} No such contribution may exist in metals in the adiabatic limit, since macroscopic fields are screened therein (Faraday-cage effect). The possible experimental manifestations of the Born charges in metals are presently under discussion.\cite{Hickox23}

In order to exploit the adiabatic HF theorem in the form of \equ{key} the parameter $\kappa$ is a component of the vector potential (as in the previous case study), while now the Hamiltonian acquires its adiabatic time dependence via the nuclear coordinates;  $\lambda$ is therefore identified with the $s$-th Cartesian coordinate. We thus get the  geometrical expressions  \[ \ZZ^*_{s,\alpha\beta} = Z_s \delta_{\alpha\beta} + \frac{\hbar c}{e} \Omega(A_\alpha,R_{s\beta}) , \label{z1}  \]   where the curvature is evaluated at $\A=0$ and $\R_s=0$.
The geometrical expression for $\ZZ^*_{s,\alpha\beta} $ is formally identical for either molecules or condensed matter. Nonetheless one has to bear in mind that the derivatives entering $\Omega(A_\alpha,R_{s\beta})$ are evaluated at  $\EEE_0=0$  in the former case and at  $\EEE=0$ in the latter case (this corresponds to impose PBCs in the perturbed $\hat{V})$. 

 \equ{z1} is well known in molecular physics since the 1980s, although obtained from rather opaque derivations;\cite{Stephens87,Amos88,Buckingham87} of course the geometrical nature of the expression was not recognized. Here instead \equ{z1} is retrieved as a simple case study of the adiabatic Hellmann-Feynman theorem, \equ{HF}. In the condensed-matter case the Born charge was first expressed as a Berry curvature in Ref. \onlinecite{King93}, although at the independent-electron level.
 
 \subsection{A sum rule: molecules, insulators, and metals}
 
 Suppose that a (neutral) molecule is rigidly translated: $\dot\R_s = \dot{\bf u}$ for any $s$. Then the linearly induced dipole vanishes, implying that \[ \sum_s \ZZ^*_{s,\alpha\beta} = 0  . \] The analogue of this for the tensors $Z^*_{s,\alpha\beta}$ in condensed matter goes under the name of ``acoustic sum rule'',\cite{PCM} but holds {\it for insulators only}. Therein, in fact, no macroscopic current flows if all the nuclei are rigidly translated at constant speed  $\dot{\bf u}$: the nuclear current and the electronic current compensate.
 
Matters are different in metals. When the nuclei are rigidly translated, {\rr a fraction of the electron density is}  ``left behind'' and some current flows: \[ j_\alpha = - \frac{e}{L^3} \left( \sum_s Z^*_{s,\alpha\beta} \right) \dot{u}_\beta . \] In the reference frame of the nuclei, the whole current is carried by the electrons which move with velocity $-\dot{\bf u}$; the current density can therefore be recast as \[ j_\alpha = e\, n^*_{\alpha\beta}\dot{u}_\beta, \] \[  n^*_{\alpha\beta} = \frac{1}{L^3}  \sum_s Z^*_{s,\alpha\beta}  = \frac{1}{L^3} [N \delta_{\alpha\beta} + {\rr \frac{\hbar c}{e}}\Omega(A_\alpha,u_\beta) \, ] ,  \label{eff} \] where $n^*_{\alpha\beta}$ assumes the meaning of effective electron density contributing to the dc current. 

The same effective electron density $n^*_{\alpha\beta} $ appears in a different phenomenon.\cite{rap165,Scalapino93,rap157} In a metal a dc macroscopic electric field $\EEE$ induces---in absence of dissipation---free acceleration of the many-electron system, whose inverse inertia is $n^*_{\alpha\beta}/m$: \[ j_\alpha(t) = \frac{e^2}{m} n^*_{\alpha\beta} \EE_\beta t . \label{free} \] The (singular) Fourier transform of \equ{free} defines the Drude term in  linear conductivity: \[ j_\alpha(\omega) = \sigma^{(\rm Drude)}_{\alpha\beta}(\omega) \EE_\beta(\omega), \]  \[ \sigma^{(\rm Drude)}_{\alpha\beta}(\omega) = D_{\alpha\beta} \left[ \delta(\omega) + \frac{i}{\pi \omega} \right] \!, \;   D_{\alpha\beta} =   \frac{\pi e^2}{m} n^*_{\alpha\beta} . \label{D} \]  The two terms in parenthesis warrant causality; the tensor $D_{\alpha\beta}$ is called  Drude weight, also known as adiabatic charge stiffness. The Dreyer-Coh-Stengel sum rule\cite{Dreyer22} follows from \eqs{eff}{D}: \[ \frac{1}{L^3} \sum_s Z^*_{s,\alpha\beta}  = \frac{m}{\pi e^2} D_{\alpha\beta} . \label{dcs} \] The message of this remarkable sum rule is that the responses to two different probes---a rigid translation of the lattice and a macroscopic electric field---sample the same material property, and are expressed by the same Berry curvature $\Omega(A_\alpha,u_\beta)$.

\section{Rotational $g$ factor and magnetizability of  a molecule} \label{sec:g}

This Section only concerns a bounded system, whose wavefunctions obey OBCs.  The rotational $g$ factor is probed by a rigid rotation of the nuclear frame, while in the magnetizability case the probe acting on the many-electron system is a macroscopic ${\bf B}$ field. I anticipate that the many-electron response to both probes is expressed by the same Berry curvature; there is therefore some analogy with the message of \equ{dcs}, where the probes instead were  a rigid lattice translation and a macroscopic $\EEE$ field.

In order to simplify the algebra, I fix here the rotation axis and the ${\bf B}$  direction, both taken as the $z$-axis.

\subsection{Rotational $g$ factor}

Suppose a molecule is rigidly rotating at velocity $\dot{\theta}$ around its center of mass; the rotational $g$ factor is defined as the ratio between the total magnetic moment $m_z$ of the molecule and its mechanical angular moment $ {\cal M}_z$, expressed in dimensionless units: \[ g = \frac{2mc}{e}\frac{m_z}{{\cal M}_z} =  \frac{2mc}{e}\frac{m_z/\dot{\theta}}{{\cal M}_z /\dot{\theta}}. \label{g} \]  The mechanical moment is \[ {\cal M}_z = \sum_s M_s R_s^2 \dot\theta ,\] where $R_s$ is the distance of the $s$-th nucleus from the axis and $M_s$ is its mass. In the following we focus solely on the numerator in \equ{g}, called the rotational magnetic moment:\cite{Arrighini67} \[ \frac{m_z}{\dot{\theta}} = \frac{e}{2c}\sum_s Z_s R^2_s   - \frac{e}{2mc}{\rr \frac{\LL_z(t)}{\dot{\theta}}}    , \label{g1} \] where {\rr $\LL_z(t)$} is the adiabatic expectation value of {\rr $\hat{\LL}_z$};  \equ{pva}.
 By adopting a gauge centered on the rotation axis one has \[ {\rr  \hat{\LL}_z} = \frac{2mc}{e} \partial_{B_z} \hat{H},  \label{gauge} \] and \equ{key} immediately yields  \[ \frac{m_z}{\dot{\theta}} = \frac{e}{2c}\sum_s Z_s R^2_s  + \hbar\, \Omega(B_z,\theta)   , \label{g2} \] 
 where the Berry curvature $\Omega(B_z,\vartheta)$ is evaluated at $\B=0$, and is $\vartheta$-independent. 
 To the best of the author's knowledge, $g$ was first expressed in terms of a Berry curvature in Ref. \onlinecite{Ceresoli02}; the form of \equ{g2} owes to Ref. \onlinecite{Stengel18}. 
 
The two terms  in \equ{g2} are the nuclear and electronic contributions to $m_z$. In view of the following, it is expedient to recast  $m_z$ as the sum of two different terms:\cite{Zabalo22} \[ m_z = m_z^{(0)} + m_z^{(1)} , \] where $m_z^{(0)}$ is the orbital moment generated by a rigid rotation of the molecular charge distribution, and  $m_z^{(1)}$ is the remaining (deformation) term. One then has:
\[ \frac{m_z^{(0)}}{\dot\theta} =  \frac{e}{2c}\left( \sum_s Z_s R^2_s - \me{\Psi_0}{\sum_i (x_i^2+y_i^2)}{\Psi_0} \right)  \label{m0} \]
\[ \frac{m_z^{(1)}}{\dot\theta} =  \frac{e}{2c} \me{\Psi_0}{\sum_i (x_i^2+y_i^2)}{\Psi_0}  + \hbar\, \Omega(B_z,\theta) . \label{max}\]

\subsection{Magnetizability}

The  linear magnetizability of  a molecule is by definition 
    \bea  \chi_{zz} &=& -\frac{\partial^2 E_0}{\partial B_z^2} =  \chi_{zz}^{(\rm dia)} +  \chi_{zz}^{(\rm para)} \label{chi} \\ &=& -\me{\Psi_0}{\frac{\partial^2 \hat{H}}{\partial B_z^2}}{\Psi_0} - 2\, \mbox{Re } \me{\Psi_0}{\partial_{B_z} \hat{H}}{\partial_{B_z} \Psi_0} \nonumber
 ,  \eea 
where the two terms are the diamagnetic and paramagnetic contributions, respectively:
\[ \chi_{zz}^{(\rm dia)} = -  \frac{e^2}{4mc^2}  \me{\Psi_0}{\sum_i (x_i^2+y_i^2)}{\Psi_0}  \]
\bea \chi_{zz}^{(\rm para)} &=& -  \frac{e}{mc}  \mbox{Re } \me{\Psi_0}{{\rr  \hat{\LL}_z} }{\partial_{B_z} \Psi_0}  \nn
&=& \frac{e^2}{2m^2c^2} \sum_{n \neq 0} \frac{\me{\Psi_0}{{\rr  \hat{\LL}_z} }{\Psi_n}\me{\Psi_n}{{\rr  \hat{\LL}_z} }{\Psi_0}}{E_n-E_0} .
\label{para} \eea
It is known since long time\cite{Arrighini67} that the rotational magnetic moment uniquely determines the paramagnetic magnetization, which therefore---in view of the above---also admits a geometrical formulation: indeed, \equ{para} is a Berry curvature in disguise.

In order to show this we observe that \[ {\rr  \hat{\LL}_z}  \ket{\Psi_0}   = - i \hbar  \ket{\partial_\theta \Psi_0} = i \hbar \sum_{n\neq 0} \ket{\Psi_n} 
\frac{\me{\Psi_n}{\partial_\theta \hat{H}}{\Psi_0}}{E_n-E_0} ; \] {\rr notice that a nondegenerate singled ground state is addressed throughout.} Replacement into \equ{para}, and using \equ{gauge}, one gets \[ \chi_{zz}^{(\rm para)} = i\frac{\hbar e}{mc}  \sum_{n \neq 0} \frac{\me{\Psi_0}{\partial_{B_z} \hat{H}}{\Psi_n}\me{\Psi_n}{\partial_\theta \hat{H}}{\Psi_0}}{(E_n-E_0)^2} . \] Finally, by comparing this to \equ{sum}, one gets \[ \chi_{zz}^{(\rm para)} = - \frac{\hbar e}{2 mc} \Omega(B_z,\theta) ; \] it is then easy to show that   \[ \chi_{zz} = -\frac{e}{2mc}\frac{m_z^{(1)}}{\dot\theta}.  \label{m1} , \] where now $\chi_{zz}$ is the {\it total} magnetizability.

\subsection{Gauge-invariance issues}

The Berry curvature $ \Omega(B_z,\theta)$ is {\it not} invariant under change of the electromagnetic gauge, because the Hamiltonian depends on $\B$ in a gauge-dependent way; in other words
$\B$ is not a ``normal'' parameter on which the Hamiltonian explicitly depends. The total magnetizability, \equ{chi}, is manifestly gauge invariant, while its two terms $\chi_{zz}^{(\rm dia)}$ and $\chi_{zz}^{(\rm para)}$ are not such. As for the rotational magnetic moment, only the term $m_z^{(1)}$ is gauge-invariant, after \equ{m1}.

All of the above results have been obtained by adopting a central gauge whose origin is on the rotation axis. We address next gauge-invariance under a restricted class of transformations, by only considering central gauges with arbitrary origin. The algebra is simpler if one keeps the gauge origin fixed and displaces the rotation axis. In fact \equ{m0} can be compactly rewritten as \[ \frac{m_z^{(0)}}{\dot\theta} = \frac{1}{2c} \intr (x^2 + y^2)\, \rho^{(0)}(\r) , \] where $ \rho^{(0)}(\r)$ is the ground-state molecular charge density (electronic and nuclear); it is then obvious that $m_z^{(0)}$ is invariant if the molecule is neutral and with vanishing static dipole.\cite{Zabalo22} Since in this case the total $m_z$ is gauge-invariant as well, the gauge transformation of $ \Omega(B_z,\theta)$ easily follows from \equ{g2}.

\section{Conclusions}

The observables addressed in this work measure the adiabatic response of either a condensed-matter system or a molecule to a slow variation of the Hamiltonian, driven by a parameter $\lambda(t)$ entering it.  The many-body setting chosen throughout allows for compact and meaningful formulas, dealing with crystalline, noncrystalline, and molecular systems on the same ground, with either interacting or noninteracting (Hartree-Fock or Kohn-Sham) electrons. 

All of the expressions presented here  are case studies of a very general adiabatic Hellmann-Feynman theorem, which provides the exact time evolution of a generic observable when the when the change of the Hamiltonian in time is made ``infinitely slow'', in the sense of Kato's theorem.\cite{Kato50} The transparent geometrical expressions displayed in this work are easily shown to be equivalent to the time-honored sum-over-states Kubo formulas of linear-response theory.

\bigskip\bigskip
\section*{Acknowledgments}
Discussions with Massimiliano Stengel on the present topics are gratefully acknowledged. Work supported by the Office of Naval Research (USA) Grant No. N00014-20-1-2847.

\section*{Data availability}
Data sharing is not applicable to this article as no new data were created or analyzed in this study.


\begin{thebibliography}{10}

\bibitem{Berry84}
{ M. V. Berry, Proc. Roy. Soc. Lond. A {\bf 392}, 45 (1984)}.

\bibitem{niu84}
{ Q. Niu and D. J. Thouless, J. Phys A {\bf 17}, 2453 (1984)}.

\bibitem{Kato50}
{ T. Kato, J. Phys. Soc. Jpn. {\bf 5}, 435 (1950)}.

\bibitem{rap_a20}
{ R. Resta, J. Phys.: Condens. Matter {\bf 12}, R107 (2000)}.

\bibitem{Vanderbilt}
{ D. Vanderbilt, {\it Berry Phases in Electronic Structure Theory} (Cambridge
  University Press, Cambridge, 2018)}.

\bibitem{Xiao10}
{ D. Xiao, M.-C. Chang, and Q. Niu, Rev. Mod. Phys. {\bf 82}, 1959 (2010)}.

\bibitem{Mead92}
{ C. A. Mead, Rev. Mod. Phys. {\bf 64}, 51 (1992)}.

\bibitem{rap168}
{ R. Resta, J. Chem. Phys. {\bf 158}, 024105 (2023)}.

\bibitem{Nafie83}
{ L. A. Nafie, J. Chem. Phys. {\bf 79}, 4950 (1983)}.

\bibitem{Patchkovskii12}
{ S. Patchkovskii, J. Chem. Phys. {\bf 137}, 084109 (2012)}.

\bibitem{Hanasaki21}
{ K Hanasaki and K. Takatsuka, J. Chem Phys. {\bf 154}, 164112 (2021)}.

\bibitem{Kohn64}
{ W. Kohn, Phys. Rev. {\bf 133}, {A171} (1964)}.

\bibitem{rap100}
{ R. Resta, Phys. Rev. Lett. {\bf 80}, 1800 (1998)}.

\bibitem{rap166}
{ R. Resta, J. Chem. Phys. {\bf 156}, 204118 (2022)}.

\bibitem{Jackson}
{ J. D. Jackson, {\it Classical Electrodynamics} (Wiley, New York, 1975)}.

\bibitem{Souza02}
{ I. Souza, J. \`I\~n\`iguez, and D. Vanderbilt, Phys. Rev. Lett. {\bf 89},
  117602 (2002)}.

\bibitem{Umari02}
{ P. Umari and A. Pasquarello, Phys. Rev. Lett. {\bf 89}, 157602 (2002)}.

\bibitem{Kittel}
{ C. Kittel, {\it Introduction to Solid State Physics}, 8th. edition (Wiley,
  Hoboken, NJ, 2005)}.

\bibitem{rap157}
{ R. Resta, J. Phys. Condens. Matter {\bf 30}, 414001 (2018)}.

\bibitem{rap165}
{ R. Resta, Phys. Rev. Research {\bf 4}, 033002 (2022)}.

\bibitem{rap_a30}
{ R. Resta, J. Phys.: Condens. Matter {\bf 22} 123201 (2010)}.

\bibitem{Dreyer22}
{ C. E. Dreyer, S. Coh, and M. Stengel, Phys. Rev. Lett. {\bf 128}, 095901
  (2022)}.

\bibitem{Huang50}
{ K. Huang and M. Born, Proc. Roy. Soc. {\bf A203}, 178 (1950)}.

\bibitem{Cochran62}
{ W. Cochran and R. A. Cowley, J. Phys. Chem. Solids {\bf 23}, 447 (1962)}.

\bibitem{Hickox23}
{ D. Hickox-Young , D. Puggioni, and J. M. Rondinelli, Phys. Rev. Materials
  {\bf 7}, 010301 (2023)}.

\bibitem{Stephens87}
{ P. J. Stephens, J. Phys. Chem. {\bf 91}, 1712 (1987)}.

\bibitem{Amos88}
{ R. D. Amos, K. J. Jalkanen, and P. J. Stephens, J. Phys. Chem. {\bf 92}, 5571
  (1988)}.

\bibitem{Buckingham87}
{ A. D. Buckingham, P. W. Fowler, and P. A. Galwas, Chem. Phys. {\bf 112}, 1
  (1987)}.

\bibitem{King93}
{ R. D. King-Smith and D. Vanderbilt, Phys. Rev. B {\bf 47}, 1651 (1993)}.

\bibitem{PCM}
{ R. Pick, M. H. Cohen, and R. M. Martin, Phys. Rev. B {\bf 1}, 910 (1970)}.

\bibitem{Scalapino93}
{ D. J. Scalapino, S. R. White, and S. C. Zhang, Phys. Rev. B {\bf 47}, 7995
  (1993)}.

\bibitem{Arrighini67}
{ G. P. Arrighini, M. Maestro, and R. Moccia, J. Chem. Phys. {\bf 49}, 882
  (1967)}.

\bibitem{Ceresoli02}
{ D. Ceresoli and E. Tosatti, Phys. Rev. Lett. {\bf 89}, 116402 (2002)}.

\bibitem{Stengel18}
{ M. Stengel and D. Vanderbilt, Phys. Rev. B {\bf 98}, 125133. (2018)}.

\bibitem{Zabalo22}
{ A. Zabalo , C. E. Dreyer, and M. Stengel, Phys. Rev. B {\bf 105}, 094305
  (2022)}.

\end{thebibliography}

\end{document}